# Pressure-induced dimerization and molecular orbitals formation in $Na_2RuO_3$ with strong correlation-enhanced spin-orbit coupling effect

Xujia Gong,[1] Wei Wang,[2] Wen Lei,[3] Feng Xiao,[1] Carmine Autieri,[4] Congling Yin,[5] and Xing Ming[1,5*]

1. College of Science, Guilin University of Technology, Guilin 541004, PR China
2. School of Materials Science and Engineering, Beihang University, Beijing 100191, PR China
3. Key Laboratory of Artificial Micro- and Nano-Structures of Ministry of Education and School of Physics and Technology, Wuhan University, Wuhan 430072, PR China
4. International Research Centre MagTop, Institute of Physics, Polish Academy of Sciences, Aleja Lotników 32/46, PL-02668 Warsaw, Poland
5. MOE Key Laboratory of New Processing Technology for Nonferrous Metal and Materials, Guangxi Key Laboratory of Optic and Electronic Materials and Devices, Guilin University of Technology, Guilin 541004, PR China

## ABSTRACT

First-principles calculations and simulations are conducted to clarify the nonmagnetic insulating ground state of the honeycomb lattice compound $Na_2RuO_3$ with $4d^4$ electronic configuration and explore the evolutions of crystal structure and electronic property under pressure. We reveal that individual Coulomb correlation or spin-orbit coupling (SOC) effect cannot reproduce the experimentally observed nonmagnetic insulating behavior of $Na_2RuO_3$, whereas the Coulomb correlation enhanced SOC interactions give rise to an unusual spin-orbital-entangled ***J*** = 0 nonmagnetic insulating state, which contrasts with the SOC assisted Mott insulating state in $d^5$ ruthenates and iridates. Furthermore, a pressure-induced structural dimerization transition has been predicted around 15-17.5 GPa. The honeycomb lattice of the high-pressure dimerized phase features with parallel pattern of the short Ru-Ru dimers aligning along the crystallographic *b*-axis direction. Accompanied with the structural dimerization, the electronic structure shows striking reconstruction by formation of molecular orbitals. Interestingly, the cooperation of Coulomb correlation together with SOC can realize a nonmagnetic insulating state in the high-pressure dimerized phase. The $d^4$ ruthenate $Na_2RuO_3$ with honeycomb lattice will provide a new platform to explore unusual physics and rich phase diagram due to the delicate interplay of lattice degree of freedom, electron correlations, and SOC interactions.

---

[*] Email: mingxing@glut.edu.cn (Xing Ming)

# I. INTRODUCTION

The competition between spin-orbit coupling (SOC), Coulomb correlations together with Hund's coupling interactions has given rise to a rich set of emergent phenomena and exotic quantum states in $4d/5d$ transition metals compounds [1-4]. Particularly, the seminal discovery of the SOC-induced Mott insulating state in the iridate $Sr_2IrO_4$ [5] has sparked considerable enthusiasm to explore unprecedented spin-orbital-entangled phases [6-8]. The spin-orbital-entangled $j_{\text{eff}}$ states become indispensable ingredients to understand the exotic electronic structures and magnetic properties of ruthenates and iridates [1-8].

Recently, the ruthenates and iridates with $d^4$ electronic configuration have attracted tremendous interest due to their rich phase diagram and potential excitonic magnetism [7,9-12]. According to the prevailing wisdom of the $j_{\text{eff}}$ picture in the strong SOC limit, the $t_{2g}$ states of the $4d$ and $5d$ orbital will split into effective $j_{\text{eff}} = 1/2$ doublet and $j_{\text{eff}} = 3/2$ quartet by SOC splitting [1,5,7]. Therefore, all the four electrons of the $d^4$ ions with $t_{2g}^4$ electronic configuration are anticipated to fully occupy the lower-energy $j_{\text{eff}} = 3/2$ states, which are separated from the higher-energy empty $j_{\text{eff}} = 1/2$ states by a gap of $3\lambda/2$ ($\lambda$ is the SOC strength). As a result, a SOC-induced $\boldsymbol{J} = 0$ nonmagnetic band insulator (often named as relativistic band insulator) should be obtained naturally. On the other hand, when the electron correlations are dominant over SOC interactions, an alternative scenario of SOC entangled $\boldsymbol{J} = 0$ nonmagnetic state has been put forward [6,7,11]. According to such a $\boldsymbol{LS}$-coupling scheme, the low-spin $t_{2g}^4$ electronic configuration shows total spin $\boldsymbol{S} = 1$ and threefold orbital degeneracy with an effective orbital moment $\boldsymbol{L} = 1$, then the intra-atomic SOC interactions give rise to a Van Vleck-type nonmagnetic $\boldsymbol{J} = 0$ singlet by antiparallel alignment of the spin and orbital angular momentum. Based on such a SOC-driven nonmagnetic $\boldsymbol{J} = 0$ ground state in $4d/5d$ compounds with $d^4$ electron configuration, theoretical predictions anticipate novel excitonic magnetism may be realized due to the condensation of spin-orbital $\boldsymbol{J} = 1$ triplons [11]. However, attributed to the delicate balance of comparable energy scales of SOC, Coulomb interaction, Hund's Coupling, and non-cubic crystal field splitting, breakdown of the $\boldsymbol{J} = 0$ nonmagnetic state and resultant complicated magnetic behavior have been extensively uncovered in real materials [7]. The $\boldsymbol{J} = 0$ non-magnetic ground state in $4d/5d$ compounds with $d^4$ electron configuration has been a controversial topic still under intense debate. For instance, conflicting magnetic behaviors for double-perovskite iridates $Sr_2YIrO_6$ and $Ba_2YIrO_6$ are reported in both experimental and theoretical studies [13-18]. The origin of these controversial magnetic properties is still ambiguous. Recent experimental works support the $J = 0$ ground state for the $Ir^{5+}$ oxides $A_2BIrO_6$ (A= Ba, Sr; B= Lu, Sc), $Sr_3NaIrO_6$ and $Sr_3AgIrO_6$, and indicate that the magnetic signals are from extrinsic sources, such as magnetic impurities or antisite disorder [19-22].

Alternately, the tetravalent ruthenates ($Ru^{4+}$, $4d^4$) $A_2RuO_3$ (A = Na, Li) [23-29] and $Ag_3LiRu_2O_6$ [30,31] feature with a nearly ideal honeycomb lattice and weaker SOC strength relative to iridates, which provides a new playground to explore unusual physics emerging from the intricate interplay of electron correlations, SOC and Hund's Coupling. A nonmagnetic state has been found in $Li_2RuO_3$ under ambient conditions, which is not related to the spin-orbital-entangled $J = 0$ state but rather associated with structural dimerization of the honeycomb lattice and molecular orbital formation in the Ru-Ru dimer [23,24]. In contrast to the distorted honeycomb structure with an armchair pattern of the long/short Ru-Ru bonds in $Li_2RuO_3$ [23-27], $Na_2RuO_3$ [28,29] and $Ag_3LiRu_2O_6$ [30,31] crystallize in a nearly perfect honeycomb lattice with almost equal Ru-Ru bond lengths, which may serve as model systems to study spin-orbital-entangled $J = 0$ physics. However, contradictory magnetism including antiferromagnetic (AFM) coupling interactions between the Ru moments and spin-orbital-entangled nonmagnetic singlet state has been observed experimentally in $Ag_3LiRu_2O_6$ at ambient pressure [30,31]. R. Kumar *et al*. studied the unconventional magnetism of $Ag_3LiRu_2O_6$, where the magnetic susceptibility $\chi(T)$ data shows a strong AFM coupling between the Ru moments without any anomaly down to 2 K, but the neutron diffraction does not detect any magnetic order down to 1.6 K [30]. In contrast, T. Takayama *et al*. performed magnetic and spectroscopic measurements, together with the quantum chemistry calculation, indicating that $Ag_3LiRu_2O_6$ is a Mott insulator and hosts a spin-orbit-entangled nonmagnetic $J = 0$ singlet [31]. In addition, conflicting measurement results of electrical transport properties and magnetic properties have also been reported for $Na_2RuO_3$. Firstly, the resistance measurements show insulating behavior, which contradicts the metallic behavior observed in the photoelectron spectroscopy measurements, where the synthesized polycrystalline samples of $Na_2RuO_3$ consisted of ~94% ordered phase and ~6% disordered phase of $Na_2RuO_3$ [32]. However, the Arrhenius fitting of resistivity data as well as x-ray absorption and emission spectra give an activation energy of 80 meV for the disordered phase of $Na_2RuO_3$, indicating very small gap [33]. In contrast, M. Tamaru *et al*. claimed that the disordered phase of $Na_2RuO_3$ shows metallic conduction [34]. Secondly, Wang *et al*. claimed that they have synthesized single crystals of $Na_2RuO_3$ and found it to be a semiconducting antiferromagnet with a Néel temperature around 30 K from bulk susceptibility and heat-capacity measurements [29]. However, it was suspected that the single crystals from their study have been mistakenly assigned to $Na_2RuO_3$, because the features of the susceptibility and heat capacity in this work are remarkably similar to those observed in $Na_3RuO_4$ [35]. In contrast, Veiga *et al*. [32] observed no sign of magnetic ordering or magnetic frustration down to 1.5 K, where the magnetic susceptibility $\chi$(T) is weakly temperature-dependent and is likely dominated by the Pauli term above 50 K. The low-temperature Curie-Weiss contribution is equivalent to approximately 10% by mass of the $S$ = 1 impurity, and the disordered phase (~6% weight percentage of the as synthesized samples) might be responsible for the magnetic impurities. Particularly, an estimated intrinsic susceptibility $\chi_0$ obtained by subtracting a Curie-like 1/T contribution is almost temperature-independent. Actually, the magnetic susceptibility of $Na_2RuO_3$ is very similar to that of $K_2RuCl_6$, another ruthenate with $d^4$ electron configuration, which features a characteristic of the Van Vleck paramagnetic susceptibility of isolated $J$ = 0 ions [36]. Furthermore, inelastic neutron scattering experiments observe nonmagnetic behavior rather than any sign of long-range magnetic ordering for $Na_2RuO_3$ [37]. The discrepancies between these previous works are perhaps originated from the different samples and the ordered or disordered polymorphs of $Na_2RuO_3$. According to previous experimental reports, $Na_2RuO_3$ crystallizes in two polymorphs, one is a disordered phase with $R\bar{3}m$ space group, another one is the ordered phase with $C2/m$ space group [28,38,39]. As a result, the x-ray diffraction patterns differ a bit that the ordered phase has diffuse scatterings arising from the honeycomb ordering in the $[Na_{1/3}Ru_{2/3}]O_2$ slabs. The ordered and disordered polymorphs exhibit significant differences in the electrochemical performances [38]. A comprehensive study of the bulk properties of single pure-phase ordered or disordered polymorphs of $Na_2RuO_3$ is highly required.

The conflicting and complicated magnetic properties in the $d^4$ ruthenates with honeycomb lattice point to an interesting possibility of realizing novel phases near the quantum critical point, which can be driven by doping, strain, or pressure [6,40]. However, $Ag_3LiRu_2O_6$ displays successive pressure-induced structural transitions and remains nonmagnetic behavior rather than excitonic magnetism [31]. Density functional theory (DFT) calculations predict a pressure-induced structural transition from the idea honeycomb lattice (space group $C2/m$) to a dimerized structure (space group $P2_1/m$) for the ordered phase $Na_2RuO_3$ at ~3 GPa [41]. In spite of the theory-predicted transition pressure of 3 GPa is easily accessible, the structural transition has not been confirmed experimentally yet. Furthermore, ignoring the electronic correlations and SOC interactions, electronic structure calculations indicate that $Na_2RuO_3$ showing metallic characteristics before and after the structural transition, which are inconsistent with recently observed nonmagnetic insulating behavior in the ordered phase $Na_2RuO_3$ [32,37].

The conflicting experimental results raise a question of whether a spin-orbital-entangled state can provide a suitable description for the nonmagnetic insulating behavior of the ordered phase $Na_2RuO_3$. Furthermore, it is imperative to reevaluate the structural stability of $Na_2RuO_3$ under pressure because previous prediction of the pressure-induced dimerized transition for the ordered phase $Na_2RuO_3$ at 3 GPa is still unverified by experiment. In the present work, we revisit the electronic structure of the ordered phase $Na_2RuO_3$ under ambient conditions by explicitly considering the Coulomb interactions and SOC, and explore the evolution of the honeycomb lattice and electronic structure as a function of applied pressure by first-principles calculations based on DFT. The calculated results indicate that only SOC interactions combined with Coulomb repulsion can give rise to the insulating nonmagnetic ground state of the ordered phase $Na_2RuO_3$. Furthermore, we predict a pressure-induced structural dimerization at 17.5 GPa, which is accompanied with electronic structure transformations and the emergence of molecular orbital. The crucial role of Coulomb correlation conspiring with SOC interactions in the honeycomb lattice ruthenate $Na_2RuO_3$ provides a good platform to study the spin-orbital-entangled physics. These theoretical results will attract further experimental confirmation of pressure-induced dimerization as well as electronic structure transition from the nonmagnetic insulating spin-orbital-entangled state to molecular orbitals state.

## II. STRUCTURE MODEL AND COMPUTATIONAL DETAILS

The crystal structures of $Na_2RuO_3$ have been characterized with an ordered phase in monoclinic space group $C2/m$ or $C2/c$ and a disordered phase in rhombohedral space group $R\bar{3}m$ [28,32,38,39]. The Na and Ru ions are randomly distributed in the $[Na_{1/3}Ru_{2/3}]O_2$ slabs in the disordered phase, whereas the ordered phase features with a honeycomb-ordered $[Na_{1/3}Ru_{2/3}]O_2$ slabs. As shown in **FIG. 1**(a), the edge-sharing $RuO_6$ octahedra form hexagons with Na ions in the cavity, and the hexagons share edge to form a quasi-two-dimensional honeycomb $[Na_{1/3}Ru_{2/3}]O_2$ slabs intercalated with Na ions in the interlayer spacing. Among the six Ru-Ru bonds of one hexagon, two (four) of them are labeled as X (Y), X and Y bonds are almost the same according to experimentally characterized lattice parameters (1−X/Y = −0.000734) (**FIG. 1**(b)). The tiny differences of the X and Y bonds lead to small deviations of the hexagon interior angles ($\alpha_1 = 120.022°$ and $\alpha_2 = 119.956°$) from the case of ideal regular hexagon of 120°. As the sister compound of $Na_2RuO_3$, $Li_2RuO_3$ crystallizes in *C*2/*m* space group above $T_c = 540$ K, which transform to *P2₁/m* space group below $T_c$ [23]. Although the space groups of *C*2/*m* and *C*2/*c* differ a bit, our calculated results indicate that their energies are almost the same (detailed lattice parameters and energies can be found in **Table S1** in the Supplemental Material [42], and the corresponding atomic positions of $Na_2RuO_3$ were tabulated in **Tables S2-S4**). In addition, we also optimized the structure with *P2₁/m* space group by substitution Li atoms in $Li_2RuO_3$ with Na [23], but it has a higher energy and a dimerized Y bond (2.695 Å). Obviously, the dimerized structure with *P2₁/m* space group is inconsistent with the non-dimerized phase of the ordered phase $Na_2RuO_3$ observed at ambient conditions. According to the most recent experimental reported results [39], the ordered monoclinic phases of $Na_2RuO_3$ with *C*2/*m* space group will be employed in further simulation process.

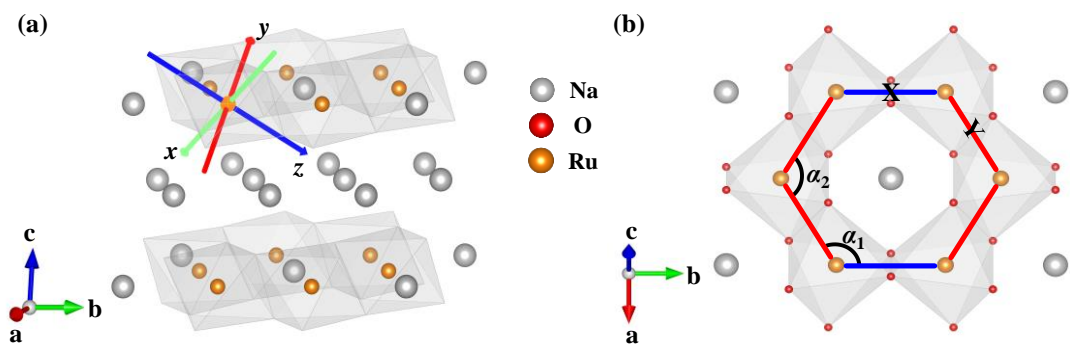

**FIG. 1.** (a) Conventional cell of the ordered phase $Na_2RuO_3$ with honeycomb-ordered $[Na_{1/3}Ru_{2/3}]O_2$ slabs. For clarity, the Na/Ru atoms are shown as grey/golden balls, while O atoms at the apexes of $RuO_6$ octahedra are not shown. We set up a local coordinate system (*x*, *y*, *z*) approximately along three perpendicular O-Ru-O bonds of a $RuO_6$ octahedron. (b) Schematic diagram of the hexagonal *ab* plane of the honeycomb $Na_2RuO_3$, together with the Ru-Ru bonds (denoted as X (blue) and Y (red) bonds) and interior angles ($\alpha_1$ and $\alpha_2$) in the hexagons formed by edge-sharing $RuO_6$ octahedra.

We performed all of the DFT calculations by employing the projector-augmented wave (PAW) method [43] as implemented in the Vienna *ab initio* simulation package (VASP) [44,45], jointing with the full-potential linearized augmented plane wave (FP-LAPW) method as implemented in the WIEN2K package [46]. The generalized gradient approximation (GGA) proposed by Perdew, Burke and Ernzerhof (PBE) was chosen to describe the exchange correlation potential [47]. On-site Coulomb interaction $U$ = 3 eV and the Hund's coupling $J_H$ = 0.3 eV (except otherwise specified) for the Ru atoms were used to deal with the electronic correlation by the DFT+$U$ method introduced by Liechtenstein *et al* [48]. The DFT+$U$ method has been widely employed to study the nonmagnetic compound with $d^4$ electronic configuration, such as the $K_2OsX_6$ (X=F, Cl, and Br) [49], $Li_2RuO_3$ [26], $SrIr_2O_6$ [50], $NaIrO_3$ [51]. Especially, the Coulomb correlation interactions ($U$) play an important role in realizing the insulating band gap of the dimerized honeycomb ruthenate $Li_2RuO_3$ features with molecular orbitals, and give a reasonable description of the experimental observed anisotropy of the susceptibility [26]. Therefore, the DFT+$U$ method is suitable to study on $Na_2RuO_3$ with $d^4$ electronic configuration. All the structural optimization was carried out by using the VASP code, where the kinetic-energy cutoff was set to 520 eV, and $k$ points spacing was set to $2\pi \times 0.2$ Å$^{-1}$, while the convergence criteria of total energy and atomic force were set to $10^{-6}$ eV and 0.01 eV/Å, respectively. To explore the stability of the honeycomb lattice under pressure, we performed simulations of the structural evolution under compression. The simulation started from the experimental atomic configuration by switching off the symmetry constraints (corresponding to the $P$1 space group) and mimicked the application pressure by gradually increasing the pressure parameter with a step of 2.5 GPa up to 40 GPa. The electronic structures including band structures and density-of-state (DOS) were cross-checked by the VASP and WIEN2K packages.

As shown in **Table S5** [42], the optimization results of the structural parameters at ambient condition within GGA+$U$ are in better agreement with the experimental data [39], the deviations of lattice parameters from the experimental results after geometry optimization are less than 1.2% at ambient conditions. In contrast, the optimized lattice parameters dramatically deviate from the experimental results when Coulomb interaction $U$ was not considered (see the comparison in **Table S5** [42]). Particularly, the X bond shows very obvious shortening (3.027 Å) with respect to the Y bonds (3.197 Å), which is obviously inconsistent with the experimental results of the non-dimerized phase at ambient condition. The agreement between the experimental data at ambient conditions and the theoretically calculated lattice parameters within GGA+$U$ method confirms the validity of the calculation parameters.

## III. RESULTS AND DISCUSSION

### A. The nonmagnetic ground state under ambient conditions

The band structures of $Na_2RuO_3$ under ambient conditions are displayed in **FIG. S1** [42], and the corresponding projection of DOS are presented in **FIG. 2**. As shown in **FIG. S1**(a) and **FIG. 2**(a), we could not open a band gap when using the standard GGA for our DFT calculation. Though there are structural distortions for the $RuO_6$ octahedra, the $4d$ orbitals of the $Ru^{4+}$ ions have been split into two-fold degenerate $e_g$ states ($d_{x^2-y^2}$ and $d_{z^2}$) and three-fold degenerate $t_{2g}$ states ($d_{xy}$, $d_{yz}$ and $d_{zx}$). The $t_{2g}$ orbitals span over the Fermi level ranging from -1.3 to 0.4 eV and result in a metallic state, which are separated from the fully empty $e_g$ states by a large crystal-field splitting (10$D$q). We expect that the insulating gap will be opened by a modest onsite Coulomb repulsion $U$ among the Ru $4d$ states. Regrettably, although the electronic states riding on the Fermi level has been split off by a pseudogap (**FIG. 2**(b)), the electronic structure still keeps a metallic characteristic by GGA + $U$ calculations with a reasonable Coulomb interaction parameter $U$ of 3 eV [52] (**FIG. S1**(b)). Particularly, even much larger $U$ value up to 5 eV is employed, it is still unable to get an insulating gap (see the band structure and corresponding DOS calculated within GGA + $U$ in **FIG. S2**(a) [42]), implying that the correlation effects are insufficient to realize the insulating behavior of $Na_2RuO_3$.

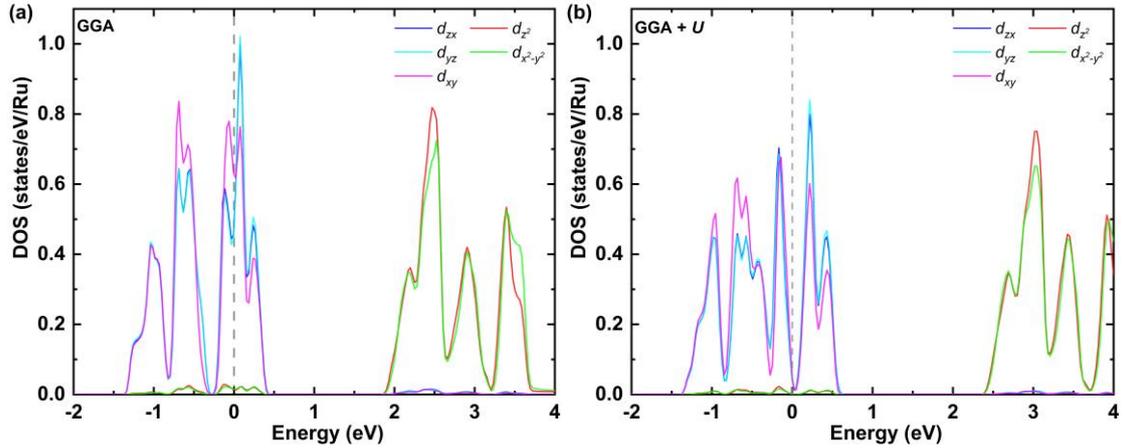

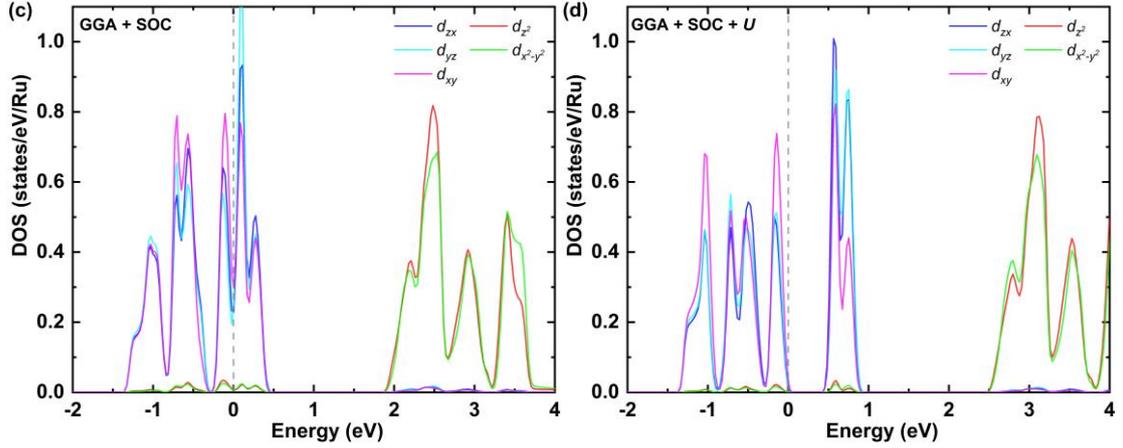

**FIG. 2.** Comparison of the projected DOS of the 4$d$ state of Ru ions for the nonmagnetic Na$_2$RuO$_3$ at ambient pressure calculated by WIEN2K code within: (a) GGA, (b) GGA + $U$, (c) GGA + SOC and (d) GGA + SOC + $U$. The Fermi level is set to 0 eV. Only spin-up results are plotted for spin-polarization GGA and GGA + $U$ calculations because the results for the spin-up and spin-down components are identical.

To further explore the underlying effects of SOC on the electronic structure of Na$_2$RuO$_3$, we consider SOC interactions by GGA + SOC calculations. At first glance, SOC interactions don't impose obvious impact on the dispersion of the band structure (see **FIG. S1** for a comparison of the band structures calculated without and with SOC [42]), and the electronic structure essentially remains the metallic characteristic (**FIG. S1**(c) and **FIG. 2**(c)). This result looks reasonable because the atomic SOC strength would be much weaker in the 4$d$ Ru atom (about 0.1 eV) as compared with 5$d$ Ir atom (about 0.4 eV) [53], therefore, the spin-orbit splitting is far less than the $t_{2g}$ bandwidth and is not strong enough to generate a band gap by SOC alone. Interestingly, although individual SOC or $U$ has relatively small influences on the dispersion of the band structures, an insulating gap of ~0.6 eV has been opened up around the Fermi level in case SOC together with $U$ are considered simultaneously (**FIG. S1**(d) and **FIG. 2**(d)). The collaborative effect of the onsite Coulomb repulsion $U$ together with SOC points towards an electron-electron correlation enhanced SOC effect [54] as proposed earlier in the 4$d$ transition metal compounds Sr$_2$RuO$_4$ [55-57] and Sr$_2$RhO$_4$ [58,59], as well as 5$d$ iridates NaIrO$_3$ [60] and $\beta$-Li$_2$IrO$_3$ [61]. The correlation-induced enhancement of SOC effect by a factor of about two has been predicted theoretically [55,56] and demonstrated experimentally for Sr$_2$RuO$_4$ [57], which plays crucial role in the electronic structure and prominently improves the theoretical description of its Fermi surface. In addition, density matrix renormalization group and multiorbital dynamical mean-field theory (DMFT) studies indicate that the interplay between Coulomb correlation effect and SOC can realize a nonmagnetic relativistic band insulator in the $t_{2g}^4$ electron system [9,10].

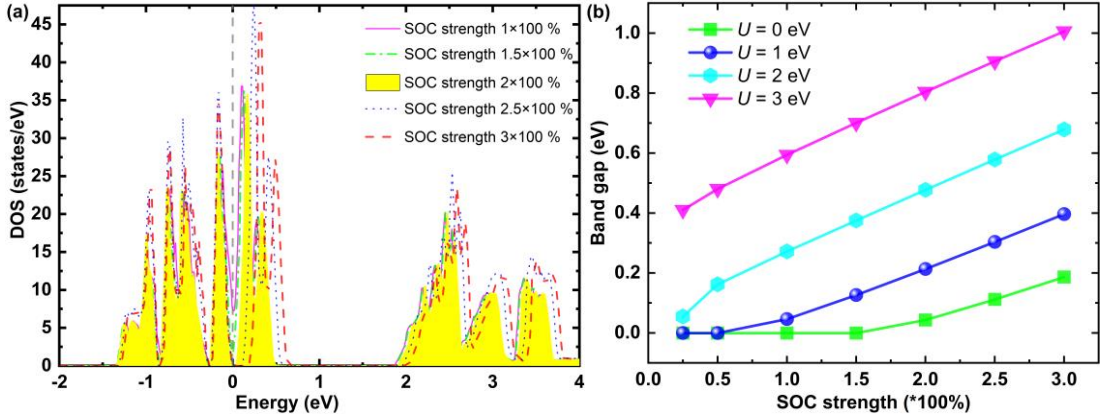

**FIG. 3.** (a) The evolutions of total DOS of Ru 4$d$ states as a function of SOC strength within GGA+SOC calculations (the opening of insulating gap by enhancing SOC strength to a factor of two has been highlighted), (b) the changes of band gaps as a function of SOC strength as well as Coulomb interaction $U$ within GGA+SOC+$U$ calculations.

The opening of the nonmagnetic insulating gap in Na$_2$RuO$_3$ can be attributed to the enhancement of SOC by correlation effects. We find that just enhance the SOC strength up to two times indeed can open the insulating gap of Na$_2$RuO$_3$ without considering the electronic correlation $U$ (**FIG. 3**(a)). In addition, the band gap value further increases linearly along with increasing SOC strength once the gap has been opened (**FIG. S3**(a) [42]), which is consistent with the anticipation that the charge gap is proportional to the SOC strength (3$\lambda$/2) [1]. On the one hand, the insulating gap can be opened by a small Coulomb interaction $U$ of 1 eV if a normal SOC strength has been taken into account, and the band gap increases gradually along with increasing $U$ (**FIG. S3**(b) [42]). On the other hand, along with the increasing Coulomb interaction $U$, the critical SOC strength to open the band gap gradually decreases from two times (bare SOC without $U$, $U$ = 0) to normal level ($U$ = 1 eV) and to 25% ($U$ = 2 eV) (**FIG. 3**(b)). Especially, the slope of band gap vs SOC strength increases obviously from 0.143 ($U$ = 0 eV) to 0.176 ($U$ = 1 eV), 0.203 ($U$ = 2 eV) and 0.216 ($U$ = 3 eV) along with the Coulomb interaction $U$ increasing from 0 eV to 3 eV, which implies the effective SOC strength $\lambda$ (derived from the slope, which is equal to 3$\lambda$/2) has been enhanced from 0.095 ($U$ = 0 eV) to 0.117 ($U$ = 1 eV), 0.135 ($U$ = 2 eV) and 0.144 ($U$ = 3 eV) eV by increasing the electron correlation (**FIG. S3**(b) [42]). According to our calculation, the effective SOC strength $\lambda$ is 0.095 eV without $U$, which has been enhanced up to 0.144 eV by about 1.5 times when $U = 3$ eV. However, without considering Coulomb correction $U$, the band gap cannot open up even if the SOC strength is enhanced to 1.5 times of the normal value. It should be noted that we are unable to adjust the absolute value of the SOC constants during the simulation, but we can manipulate the relative strength of the SOC, which has been demonstrated to be an effective method to study the $J = 0$ nonmagnetic insulating state of $K_2OsX_6$ (X = F, Cl, and Br) with $5d^4$ electronic configuration [49].

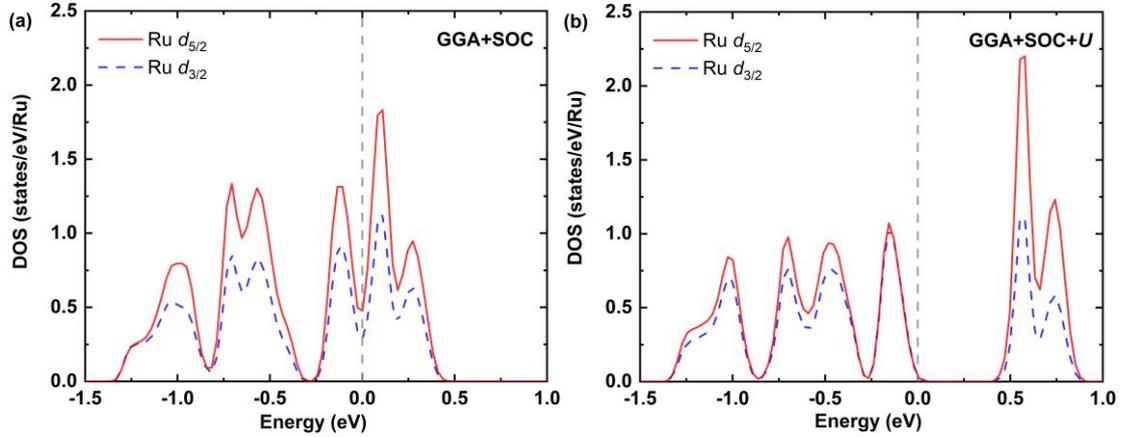

**FIG. 4.** The Ru 4$d$ DOS projected onto the fully relativistic basis set of Ru $d_{3/2}$ (blue) and $d_{5/2}$ (red) states for $Na_2RuO_3$ within (a) GGA+SOC and (b) GGA+SOC+$U$ ($U = 3$ eV).

We have demonstrated that the SOC interactions play a huge impact on the electronic structure of $Na_2RuO_3$. Particularly, the combined interplay of electronic correlation with SOC effect has induced the metal-insulator transition (MIT) and nonmagnetic state, which are consistent well with recently experimental results of $Na_2RuO_3$ [32,37]. However, as shown in **FIG. S1** [42], the individual impact of the SOC effect imposed on the band dispersion of $Na_2RuO_3$ is not as huge as observed in 5$d$ iridates such as $Sr_3M IrO_6$ ($M$ = Sr, Na, Li) [21-22], $\beta$-$Li_2IrO_3$ [61,62] and $Ca_4IrO_6$ [63], that the SOC interactions are strong enough to cause obvious separation between the $j_{eff} = 1/2$ and $j_{eff} = 3/2$ states in these iridates. In order to get further insight to the evolution of the electronic structure of $Na_2RuO_3$ along with the Coulomb enhanced SOC effect, the projection of Ru 4$d$ states onto the fully relativistic basis set of $d_{3/2}$ and $d_{5/2}$ states is presented in **FIG. 4**. The $j_{eff}$ states can be expressed in terms of $|j, j_z\rangle$ ($j$ = 5/2, 3/2) basis (the detailed derivation can be found in the Supplemental Material [42]), the $j_{eff} = 1/2$ state consists of pure $d_{5/2}$ character and the $j_{eff} = 3/2$ state has both $d_{5/2}$ and $d_{3/2}$ components in an ideally octahedral environment [3,62,64,65]. However, without considering the Coulomb correlation, the $j_{eff} = 1/2$ and $j_{eff} = 3/2$ states completely mix with each other around the Fermi level, which originates from the weak SOC strength in $Na_2RuO_3$ together with the octahedral distortion and complicated structural connectivity of the honeycomb lattice [21,65,66]. By contrast, the effective SOC effect is enhanced by Coulomb correlation, the $j_{eff} = 1/2$ and $j_{eff} = 3/2$ states are separated by an insulating gap, and the $j_{eff} = 1/2$ state above Fermi level becomes more dominant while $j_{eff} = 3/2$ component decreases, providing circumstantial evidence in favor of the $j_{eff}$ picture to describe the electronic structure of $Na_2RuO_3$. The Coulomb correlation enhanced SOC effect and resultant MIT in $Na_2RuO_3$ like the case of iridates $NaIrO_3$ with $5d^4$ electronic configuration, where the correlation interaction enhances the SOC effect, inducing a band insulating phase with renormalized band structure [60]. In addition, the mechanism of the MIT in $Na_2RuO_3$ is completely distinct from the SOC assisted Mott MIT in ruthenates and iridates with $d^5$ electronic configuration, in which the strong SOC effects firstly produce a half-filled $j_{eff} = 1/2$ states, then the individual Coulomb correlation interactions open a band gap in the $j_{eff} = 1/2$ states [5].

We remark that the intrinsic nature of the nonmagnetic insulating behavior of $Na_2RuO_3$ may be different from other insulating $d^4$ ruthenates such as the honeycomb lattice $Ag_3LiRu_2O_6$ [31] and cubic phase $K_2RuCl_6$ [36], in which SOC plays a vital role in the formation of SOC entangled $\boldsymbol{J} = 0$ nonmagnetic singlet according to a $\boldsymbol{LS}$-coupling scheme [6,7,11]. This spin-orbit entanglement has been proposed as a generic characteristic of the $d^4$ ruthenates [36], in which both spin and orbital moments should be survived for the $\boldsymbol{LS}$-coupling induced $\boldsymbol{J} = 0$ nonmagnetic state. Our calculated results indicate that the spin and orbital moments have completely quenched for the nonmagnetic $Na_2RuO_3$, which is inconsistent with the $\boldsymbol{LS}$-coupling scenario. Alternatively, an itinerant quasi molecular orbitals (QMOs) picture has been proposed to describe the nonrelativistic electronic structure of $A_2IrO_3$ (A = Li, Na) [67-69], $RuCl_3$ [70] and $Ag_3LiIr_2O_6$ [71]. These compounds feature with $d^5$ electronic configuration, honeycomb backbone and oxygen assisted $d$-$d$ hybridizations. SOC effect can destroy the QMO and leads to the formation of the relativistic $j_{eff}$ states. Interestingly, the much smaller SOC effect in $RuCl_3$ almost plays no impact on the QMOs, but the combination of Coulomb correlation and SOC effect dramatically changes the electronic structure of $RuCl_3$, once again providing a hint of the correlation enhanced SOC effect and resultant $j_{eff}$ states [53,70]. Therefore, a low-energy description in terms of relativistic $j_{eff}$ picture may be still valid for $Na_2RuO_3$, and the $\boldsymbol{J} = 0$ nonmagnetic insulating state constructs a basis of the excitonic picture remains to be confirmed experimentally.

## B.  Structural dimerization under hydrostatic pressure

In order to inspect whether the hexagonal honeycomb building blocks of $Na_2RuO_3$ is susceptible to pressure, we simulate the structural stability by applying hydrostatic pressure up to 40 GPa. The evolutions of the crystal structure under compression are summarized in **FIG. 5** and **Tables S6-S8** in the Supplemental Material [42]. As shown in **FIG. 5**(a), the lattice constants *a*, *b*, and *c* are compressed gradually along with increasing hydrostatic pressure up to 15 GPa, which results in a successive contraction of the cell volume (**FIG. 5**(b)). The response of out-of-plane lattice constant *c* is greater than those of the in-plane lattice constants *a* and *b* (rates of change are $\Delta a$ ~3.02%, $\Delta b$ ~3.32%, and $\Delta c$ ~5.9%, respectively), which is consistent with the layered structure characteristics of $Na_2RuO_3$. The lattice parameters undergo remarkable changes once the pressure increases to 17.5 GPa, implying a structural transformation occurs at 15-17.5 GPa. Around 15-17.5 GPa, the lattice constant *a* only slightly changes, but the lattice constant *b* experiences a sudden reduction accompanied with an upturn for the lattice constant *c*, which gives rise in significant shrinkage of volume and monoclinic lattice angle $\beta$ (**FIG. 5**(b)). Further inspection of the structural detail reveals that the Ru-Ru bond lengths (denoted as X and Y bonds) are almost the same under ambient conditions, which monotonically decrease with pressure, but they change significantly when the pressure exceeds 15 GPa (**FIG. 5**(c)). As shown in **FIG. S4** [42], the Y bond is elongated a bit, whereas the X bond shows an obvious dimerization from 3.00 Å to 2.61 Å, leading to a length difference of about 0.4 Å between the X and Y bonds, the Ru-O bond lengths and O-Ru-O bond angles in the $RuO_6$ octahedra also change apparently. The structural transition is originated from the dimerization of the Ru-Ru bonds (X bond) along the *b* axis direction, which is in line with the abrupt shrinking of the in-plane lattice constant *b*. The bond lengths of the Ru-Ru dimers (2.61 Å) are very close to those of $Li_2RuO_3$ (2.57 Å) [23] and $Ag_3LiRu_2O_6$ (2.51 Å) [31]. Accompanied by the dimerization of the Ru-Ru bonds, the hexagon of structural units severely deviates from the ideal honeycomb, resulting in distinctly different interior angles (**FIG. 5**(d)).

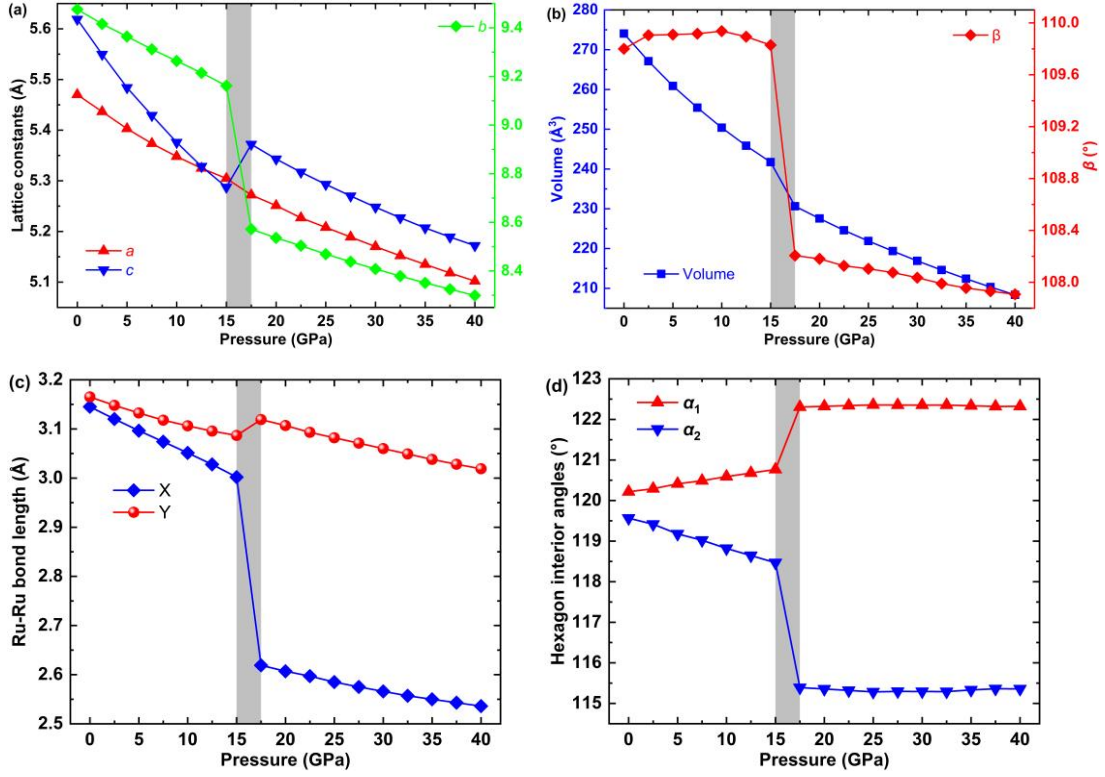

**FIG. 5.** Evolutions of the crystal structure of the ordered phase $Na_2RuO_3$ under compression: (a) lattice constants, (b) unit cell volume and the $\beta$ angle of the monoclinic lattice, (c) Ru-Ru bond lengths, (d) hexagon interior angles. The shadow areas indicate the structural transitions. The data points above 15 GPa are obtained by imposing the $C2/m$ space group.

By a detailed inspection of the structure data at pressure of 17.5 GPa as tabulated in **Table S6** in the Supplemental Material [42], we find there is a tiny difference for the Y type Ru-Ru bond lengths, and the hexagon interior angles $\alpha_1$ and $\alpha_2$ also show very small differences, respectively. Further structural analysis using the FINDSYM program [72] demonstrates that the new dimerized structure at 17.5 GPa can be assigned to either monoclinic $C2/m$ or $P2_1/m$ space group, depending on the tolerance. The detailed lattice parameters and enthalpies of the high-pressure dimerized phase by imposing these two different space groups are tabulated in **Tables S7** and **S8** in the Supplemental Material [42]. The calculated equilibrium lattice parameters of the conventional cell are $a = 5.272$ Å, $b = 8.572$ Å, $c = 5.372$ Å, and $\alpha = \gamma = 90°$, $\beta = 108.206°$ for the dimerized monoclinic $C2/m$ phase at 17.5 GPa. By comparison, those of the monoclinic $P2_1/m$ phase are $a = 5.266$ Å, $b = 8.569$ Å, $c = 5.379$ Å, and $\alpha = \gamma = 90°$, $\beta = 108.165°$, respectively. The two structures show very small differences. As shown in **FIG. 6**, the Y type Ru-Ru bonds are unequal in the $P2_1/m$ phase, and the hexagon interior angles deviate from the ideal 120° with three different values. Despite the symmetries are different for these two monoclinic structures, the dimerized patterns are the same, which can be named as parallel type [73]. Furthermore, the *P*2$_1$/*m* structure is closely related to the *C*2/*m* structure, the monoclinic *P*2$_1$/*m* space group is one of the subgroups of *C*2/*m* space group. As shown in **FIG. 6**(c), the simulated x-ray diffraction patterns of the *C*2/*m* and *P*2$_1$/*m* structures are almost identical, indicating it is difficult to distinguish these two high-pressure structures. Further increasing the pressure up to 40 GPa, all lattice parameters change regularly with pressure, which indicates that the dimerized phase of Na$_2$RuO$_3$ has been stabilized. The calculated enthalpy of the high-pressure *C*2/*m* phase is always comparable to that of the *P*2$_1$/*m* one. The enthalpy differences between *C*2/*m* and *P*2$_1$/*m* phases are less than 0.25 meV/atom (**FIG. 6**(d)), which may be within the calculation precision of the VASP code. Therefore, we cannot definitely identify the high-pressure structures of Na$_2$RuO$_3$ at present, and hereafter we maimly present the results of the high-pressure dimerized phase with *C*2/*m* space group in the main text due to its higher symmetry with respect to the *P*2$_1$/*m* phase.

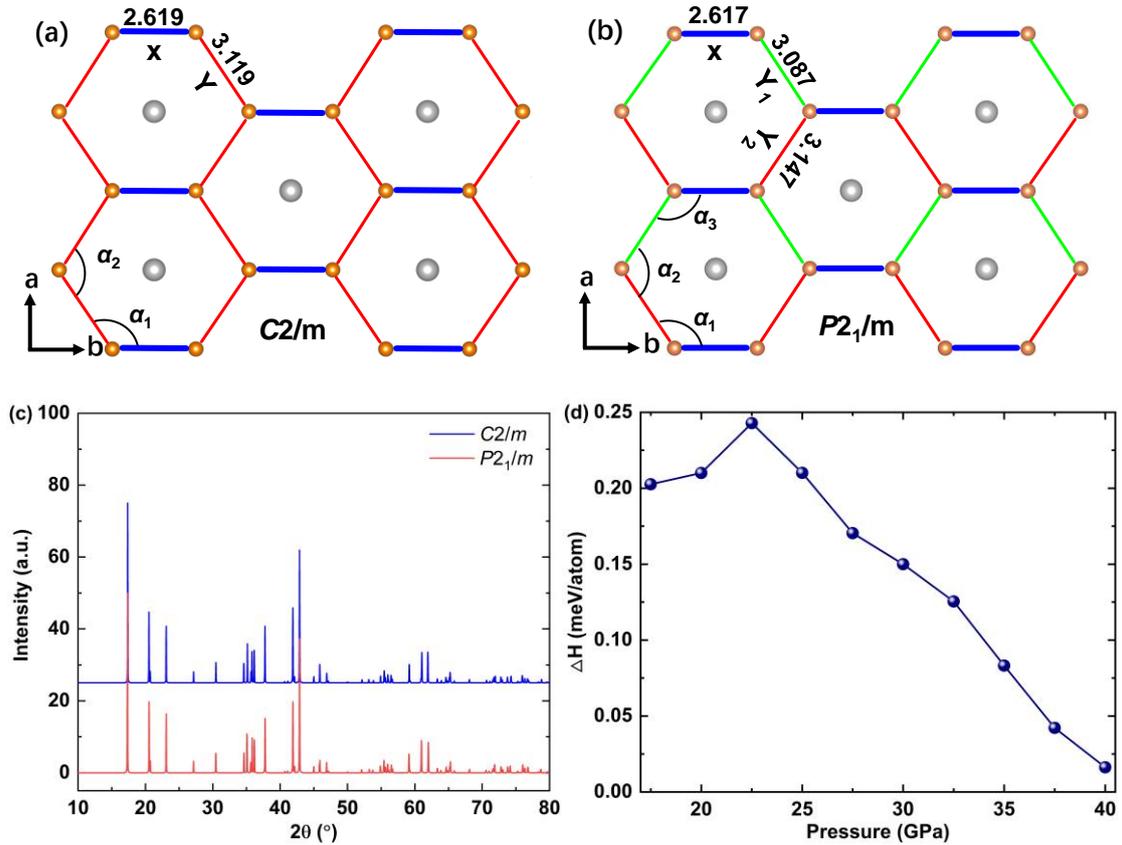

**FIG. 6.** Schematic diagram of the dimerization patterns of the Ru-Ru bonds in the hexagon formed by edge-sharing RuO$_6$ octahedra for the crystal structure at 17.5 GPa within (a) *C*2/*m* and (b) *P*2$_1$/*m* space group. (c) Simulated x-ray diffraction patterns ($\lambda = 1.541$ Å) of the *C*2/*m* and *P*2$_1$/*m* phases

at 17.5 GPa, which are handled by the powder diffraction tool within the Reflex module in Materials Studio program package. (d) The enthalpy differences (ΔH) between the dimerized $C2/m$ and $P2_1/m$ phases along with the increasing pressure.

The lattice structure of the high-pressure dimerized $C2/m$ phase retains the ambient-pressure monoclinic symmetry of $Na_2RuO_3$, which is similar to the case of pressure-induced dimerization in $Li_2RhO_3$ [73]. In contrast, the structure symmetry of the high-pressure dimerized $P2_1/m$ phase of $Na_2RuO_3$ similar to the case of $Li_2RuO_3$, in which the space group has transformed from $C2/m$ to lower-symmetry dimerized $P2_1/m$ along with decreasing temperature ($T_c$ ~550 K) at ambient pressure [23,25]. Although the structure symmetry shows similarity between $Na_2RuO_3$, $Li_2RhO_3$ and $Li_2RuO_3$, the parallel type of dimerization pattern in $Na_2RuO_3$ differs from the armchair type of dimerization in $Li_2RuO_3$ induced by temperature decreasing or pressure-induced zigzag chains dimerization in $Li_2RhO_3$ [23-27, 73]. Different compressive behavior of honeycomb iridates $A_2IrO_3$ (A = Na, Li) under pressure has also been uncovered, structural phase transition into a dimerized triclinic phase in $Li_2IrO_3$ is observed under low pressure, while the $C2/m$ crystal structure of $Na_2IrO_3$ survive up to very high pressure without dimerization [74-79]. It could speculate that the ionic radii of the buffer elements A in the honeycomb lattice play a significant role in controlling the structure dimerization of $A_2IrO_3$ (A = Na, Li) [78,79]. In this context, it is particularly instructive to revisit the compressive behavior of $Na_2RuO_3$, the buffer elements also feature distinct ionic radii between $Na_2RuO_3$ and $Li_2RuO_3$. The Na ions occupying the center of the honeycomb plane have larger radius than Li ions, which might realize different dimerization behavior between $Na_2RuO_3$ and $Li_2RuO_3$. Experiments demonstrate that substitution 5% Na for Li has dramatically changed the dimerization pattern of $Li_2RuO_3$ from the armchair type to zigzag type [27,73]), which can be understood with chemical pressure effects by replacing Li ions with larger-radius Na ions [74]. Various types of pressure-driven dimerization in hexagonal honeycomb systems have been found [73], which are extensively studied in layered honeycomb iridates $α$-$Li_2IrO_3$ [76-79], ruthenate $Ag_3LiRu_2O_6$ [31] and $α$-$RuCl_3$ [80-83], respectively. Previous theoretical work predicts that pressure can induce dimerization transition from the monoclinic $C2/m$ phase to a lower-symmetry monoclinic $P2_1/m$ phase for $Na_2RuO_3$, the predicted armchair-type dimerization pattern and structure symmetry of the high-pressure phase of $Na_2RuO_3$ is the same as those of $Li_2RuO_3$ [41], and the critical pressure of 3 GPa is predicted from the pressure dependence on the volume, which is based on two independent simulations of the non-dimerized phase and the dimerized phase under compression. However, the structural transition has not been confirmed despite the theory-predicted transition pressure of 3 GPa is easy to implement experimentally. Comparatively, the ambient-condition structure is allowed to freely relax and the structural transition from the non-dimerized phase to the dimerized phase is realized along with the gradual loading of pressure in the present work, which is more like a real high-pressure experiment. A higher transition pressure of 17.5 GPa is predicted in the present work, and the parallel-type dimerized pattern totally differs from the armchair type of dimerization as predicted in previous work [41]. The difference might originate from whether the Coulomb interaction $U$ has been considered, the structural optimization results at ambient condition within GGA+$U$ are in good agreement with the experimental data, whereas the optimized lattice parameters dramatically deviate from the experimental results when Coulomb interaction $U$ was not considered (see the comparison in **Table S5** in the Supplemental Material [42]). Particularly, the calculation without considering Coulomb interaction $U$ has output obvious short X bonds with respect to the Y bonds (3.027 vs 3.197 Å) and deformed hexagon interior angle. Therefore, previous work may underestimate the transition pressure of the structural dimerization. In view of the higher transition pressure predicted in the present work, it deserves to apply much higher pressure to explore the structural stability of $Na_2RuO_3$. In addition, recent high-pressure experiment reveals two successive transitions in the $4d^4$ honeycomb ruthenate $Ag_3LiRu_2O_6$, where the crystal structure of the intermediate phase exhibits the same structure with that at ambient pressure, but the detailed dimerized structure of the higher-pressure phase is still unclear [31]. High quality single crystals and detailed structural characterization are desired to further clarify the complicated dimerization pattern of these honeycomb ruthenates under pressure.

### C. Electronic structure transitions under hydrostatic pressure

Previous studies indicate that the structural dimerization in the hexagonal honeycomb materials can lead to a major change of the electronic structure [23-25,31,76-83]. Firstly, we recall the fundamental electronic structure at ambient pressure in **FIG. 2** and the critical role played by the electron-correlation enhanced SOC effect in the nonmagnetic insulating ground state of $Na_2RuO_3$. Along with the pressure increases to 15 GPa, though the band gap has reduced due to the band-width broadening, the essential characteristics of the electronic structure maintain unchanged and the synergistic effects of correlation interactions and SOC still take crucial role in the electronic structure (**FIG. S5** [42]). Once the pressure increases to 17.5 GPa, the structural dimerization is accompanied with a prominent reconstruction of the electronic structure. As shown in **FIG. 7** and **FIG. S6** [42], the degeneracy of the Ru $t_{2g}$ states in the vicinity of the Fermi level has been completely removed for the high-pressure dimerized phases. The $RuO_6$ octahedra share edges to form a honeycomb structure, the bond lengths of the dimerized Ru-Ru bonds (X bonds) in $Na_2RuO_3$ (~2.62 Å) are shorter than those in Ru metal (~2.7 Å), therefore the overlap between the orbitals becomes larger when the pressure brings the two adjacent Ru ions close enough. Like other dimerized phases of $Li_2RuO_3$, $α$-$Li_2IrO_3$ and $α$-$RuCl_3$, the isolated $d_{xy}$ orbitals located in the lowest and highest energy positions of the $t_{2g}$ states, which are separated from the degenerated $d_{yz}$ and $d_{zx}$ orbitals, manifesting strong covalent interactions in the Ru-Ru dimers and the formation of bonding and antibonding states [23,24,76,77,82,83].

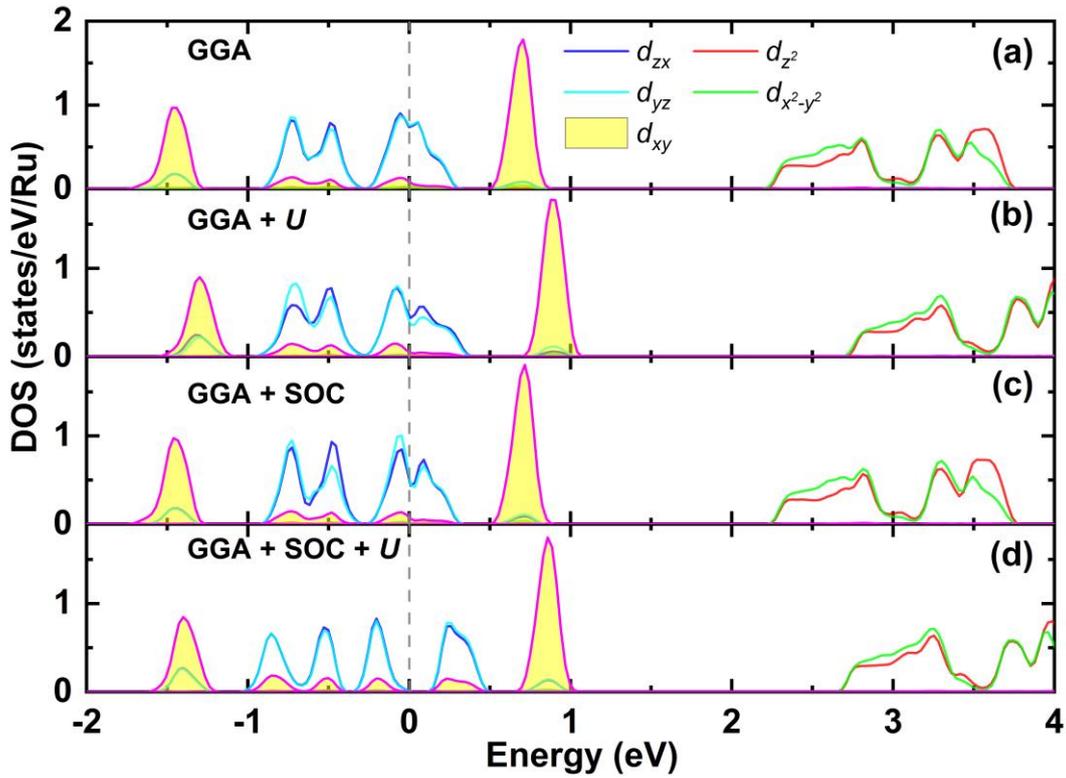

**FIG. 7.** Projected DOS of the 4$d$ state of Ru ions for the nonmagnetic dimerized $Na_2RuO_3$ at 17.5 GPa with $C2/m$ space group calculated by WIEN2K code within: (a) GGA, (b) GGA + $U$, (c) GGA + SOC and (d) GGA + SOC + $U$. The contributions of the $d_{xy}$ orbitals to the $t_{2g}$ states are highlighted by the shade region. The Fermi level is set to 0 eV. Only spin-up results are plotted for the spin-polarization GGA and GGA + $U$ calculations because the results for the spin-up and spin-down components are identical.

The electronic structure of the dimerized phase can be further analyzed by the real-space wavefunctions in the Ru-Ru dimers. As shown in **FIG. 8**, the $d_{xy}$ orbitals of the two adjacent edge-sharing RuO$_6$ octahedra form strong bonding and antibonding molecular orbitals (denoted as $\sigma$ and $\sigma^*$) due to the strong enhancement of the direct Ru-Ru hopping path along the dimerized bonds [23,24,77,83-85]. Furthermore, the rest of $t_{2g}$ states, $d_{xz}$ and $d_{yz}$ orbitals also form $\pi$ and $\delta$ types of molecular orbitals in the Ru-Ru dimers. According to this molecular orbital picture, the $\sigma$ molecular orbitals solely originate from the $d_{xy}$ orbitals, whereas the $\pi$ and $\delta$ types of molecular orbitals consist of linear combinations of $d_{xz}+d_{yz}$ and $d_{xz}-d_{yz}$ orbitals, respectively [24,25,85]. As shown in **FIG. 8** and **FIG. S7** [42], the real-space wavefunctions of the Ru-Ru dimers provide direct evidences of the formation of molecular orbital in the dimerized Na$_2$RuO$_3$. But the bonding and antibonding splitting of the $\pi$ and $\delta$ molecular orbitals are significantly weaker than the $\sigma$ molecular orbitals, which leads to the broadening of the $\pi$ and $\delta$ states and renders the two states quasi degenerate (**FIG. 7**). Ideally, the eight electrons of the two dimerized Ru$^{4+}$ ions can realize a nonmagnetic electron configuration by filling $\sigma$, $\pi$ and $\delta$ bonding orbitals and $\delta^*$ antibonding orbitals (considering the spin degree of freedom, each orbital is two-fold degenerate) [23,24,85]. However, the $\pi$ and $\delta$ bonding or antibonding states are slightly split from each other due to the smaller energy difference between them, which obscures the gap due to the band broadening and causes the Fermi level to cut through the doubly degenerate $\pi^*/\delta^*$ antibonding pair as schematic shown in **FIG. 9** [84]. Therefore, the dimerized phase shows a metallic electronic structure even at the level of DFT+$U$ (**FIG. 7**(b)). Similar to the case of ambient-pressure non-dimerized phase, an increase of the correlation strength up to 5 eV is still unable to open a gap in the degenerate $\pi^*/\delta^*$ antibonding states (**FIG. S2**(b) [42]). In contrast, first-principles DFT calculations and a combination of DFT with the cluster extension of DMFT calculations imply the Coulomb correlation $U$ plays an important role in realizing the insulating state of the low-temperature dimerized phase Li$_2$RuO$_3$ [26,86]. Intriguingly, by simultaneously including both Coulomb correlation $U$ and SOC, the degenerated $d_{yz}$ and $d_{zx}$ bands transform into several isolated narrow bands, and an insulating gap has been opened between the $\pi^*/\delta^*$ antibonding states. The MIT in the high-pressure dimerized phase once again implies the crucially collaborative effects of Coulomb correlation $U$ and SOC in Na$_2$RuO$_3$.

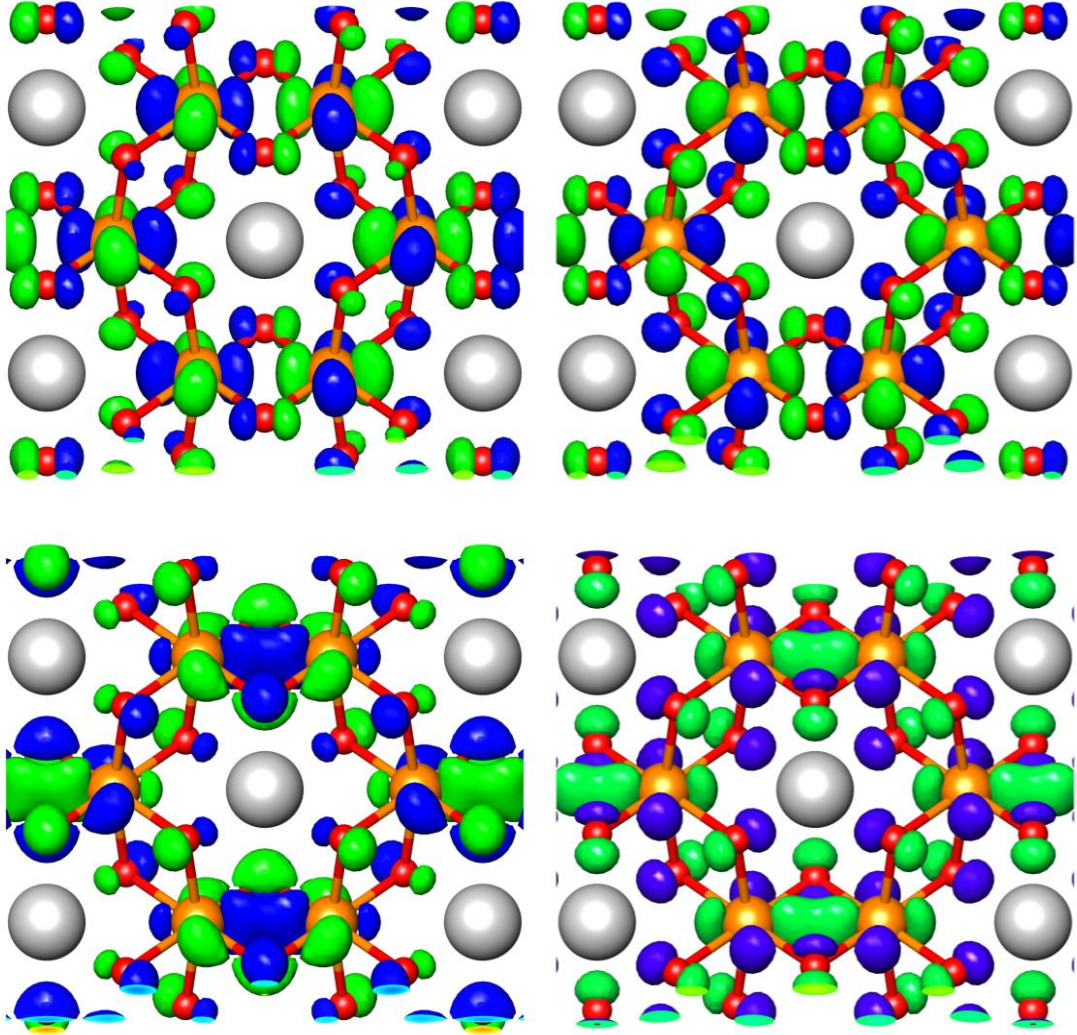

**FIG. 8.** Real-space visualization of the bonding $\sigma$ (bottom) and antibonding $\sigma^*$ (top) molecular orbitals on the Ru-Ru dimers for the nonmagnetic dimerized $Na_2RuO_3$ at 17.5 GPa with $C2/m$ space group using the VESTA tool [87], the green and blue colors of the isosurfaces denote the positive and negative wave function values, respectively.

Finally, we'd like to summarize the differences between $Na_2RuO_3$ and $Li_2RuO_3$. Firstly, the dimerization of Ru-Ru bons in $Na_2RuO_3$ is driven by applied pressure, whereas the dimerization in $Li_2RuO_3$ is due to temperature decreasing. Secondly, the dimerized phases of $Na_2RuO_3$ and $Li_2RuO_3$ show distinct dimerization patterns of parallel and armchair types [23-25]. Thirdly, the Coulomb correlation $U$ and SOC interactions play different role in realizing the insulating state of these ruthenates, while the Coulomb correlation $U$ can open the insulating gap of the dimerized phase $Li_2RuO_3$ [26,86], whereas only a combination of Coulomb correlation $U$ and SOC effect can generate the insulating gap of both the ambient-pressure non-dimerized normal phase and the high-pressure dimerized phase of $Na_2RuO_3$. In addition, a combination of x-ray diffraction and high-energy x-ray measurements indicate that the dimerized Ru-Ru dimers seem to survive locally in $Li_2RuO_3$ above the transition temperature, and the high-temperature phase serves as an example of valence bond liquid phase showing orbital-selective metallicity [25,85]. Therefore, it is interesting to clarify the structural details and electronic properties of $Na_2RuO_3$ under high pressure in further experiments.

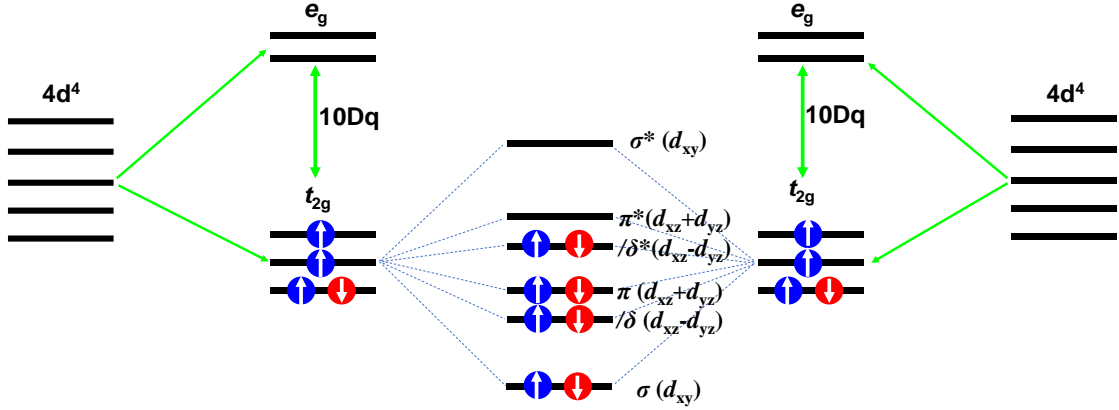

**FIG. 9.** Schematic diagram of the arrangements of the molecular orbitals and electronic configuration of the Ru-Ru dimers in the dimerized phase $Na_2RuO_3$. The $d$ orbitals have been split into two-fold degenerate $e_g$ states and three-fold degenerate $t_{2g}$ states by the octahedral crystal field strength $10D$q. The strongest bonding $\sigma$ and antibonding $\sigma^*$ molecular orbitals occupy the lowest and highest energy levels, while the splitting in the bonding $\pi/\delta$ (antibonding $\pi^*/\delta^*$) molecular orbitals is not strong enough, giving rise to a mixed state. The arrows directions denote the spin-up and spin-down states.

## IV. CONCLUSIONS

In conclusion, we have studied the electronic structure and structural stability of the $d^4$ ruthenate $Na_2RuO_3$ by first-principles calculations. The calculated results indicate that the insulating nonmagnetic $J = 0$ ground state of $Na_2RuO_3$ at ambient conditions is originated from the electron correlation enhanced SOC effect. Furthermore, we discover a pressure-induced structural dimerization of the Ru-Ru bonds featuring with a parallel pattern of the Ru-Ru dimers, which leads to an electronic structure reconstruction by emergence of molecular orbital. Interestingly, Coulomb interactions collaborating with SOC effect can promote the band-gap opening for the nonmagnetic high-pressure dimerized phase. The crucial role played by the SOC effect in realizing the nonmagnetic insulating behavior of the ambient-pressure non-dimerized phase and high-pressure dimerized phase of $Na_2RuO_3$ shed new light on the spin-orbital physics, which deserves further theoretical and experimental investigations.

## ACKNOWLEDGMENTS

We thank Dr. G. Khaliullin and T. Takayama for fruitful discussions and help. The work was sponsored by the National Natural Science Foundation of China (No. 11864008, 12264011) and Guangxi Natural Science Foundation (No. 2018AD19200 and 2019GXNSFGA245006). C. A. is supported by the Foundation for Polish Science through the International Research Agendas Programme co-financed by EU within the Smart Growth Operational Programme (Grant No. MAB/2017/1). The computational work in this research was carried out at Shanxi Supercomputing Center, and the calculations were performed on TianHe-2.